\documentclass[useAMS,usenatbib]{mn2e}
\usepackage{graphicx}

\title[]{Near\--Infrared photometry of four metal\--rich Bulge 
globular clusters: NGC~6304, NGC~6569, NGC~6637, NGC~6638 
\thanks{Based on data taken at the ESO/NTT Telescope,
	within the observing programme 73.D-0313.}}

\author[Valenti, Origlia \& Ferraro]{E. Valenti$^{1,2}$, L. Origlia$^2$, 
        F.~R. Ferraro$^1$    \\
 $^1$ Dipartimento di Astronomia, Universit\`a di Bologna,  
      Via Ranzani 1, I-40127 Bologna, Italy ,\\
      e-mail elena.valenti3@unibo.it, francesco.ferraro3@unibo.it\\
$^2$ INAF-Osservatorio Astronomico di Bologna, Via Ranzani 1, I-40127 Bologna,
      Italy, \\
      e-mail livia.origlia@bo.astro.it \\
       }
\date{\today}

\begin{document}
\pagerange{\pageref{firstpage}--\pageref{lastpage}} \pubyear{2004}
\maketitle
\label{firstpage}

\begin{abstract}

We present high\--quality near\--Infrared photometry
of four Bulge metal\--rich globular clusters, namely:
NGC~6304, NGC~6569, NGC~6637 and NGC~6638. 
By using the observed Colour\--Magnitude Diagrams we derived a
photometric estimates of the cluster reddening and distance.
We performed a detailed analysis of the Red Giant Branch, presenting 
a complete description of morphologic parameters and evolutionary 
features (Bump and Tip).
Photometric estimates of the cluster metallicity have been obtained by
using the updated set of relations (published by our group) 
linking the metal abundance to a variety of near\--Infrared 
indices measured along the Red Giant Branch. 
The detection of the Red Giant Branch Bump and the Tip is also presented and
briefly discussed.
\end{abstract}


\section{Introduction}
Bulge globular clusters (GCs) are key templates of simple stellar 
populations to study the 
stellar and chemical evolution in the high metallicity domain.

A few optical and near\--Infrared (IR) ground\--based photometric studies of the
Bulge GCs have been performed in the past decade \citep[see e.g.][]{fkt95, 
obb96a, obb96b, guarnieri, hr99, dav92, yaz03}. 
However, the relative faintness of these targets in the optical, the
 lack of complete and homogeneous surveys, the
 modest performances of the previous generation of IR arrays
 (i.e. limited size, large fraction of bad pixels, etc.),
prevented a detailed and quantitative characterization 
of the Post Main
Sequence (Post\--MS) evolution in the high metallicity domain, as potentially 
traced by
this GC sub\--system. 
In this respect, the Two\--Micron All\--Sky Survey (2MASS) could represent a
step forward but its modest spatial resolution and photometric 
deepness prevent to sample the cluster population 
with the sufficient accuracy and statistical significance, 
particularly in the core region.
%
\begin{figure*}
\includegraphics[width=18cm]{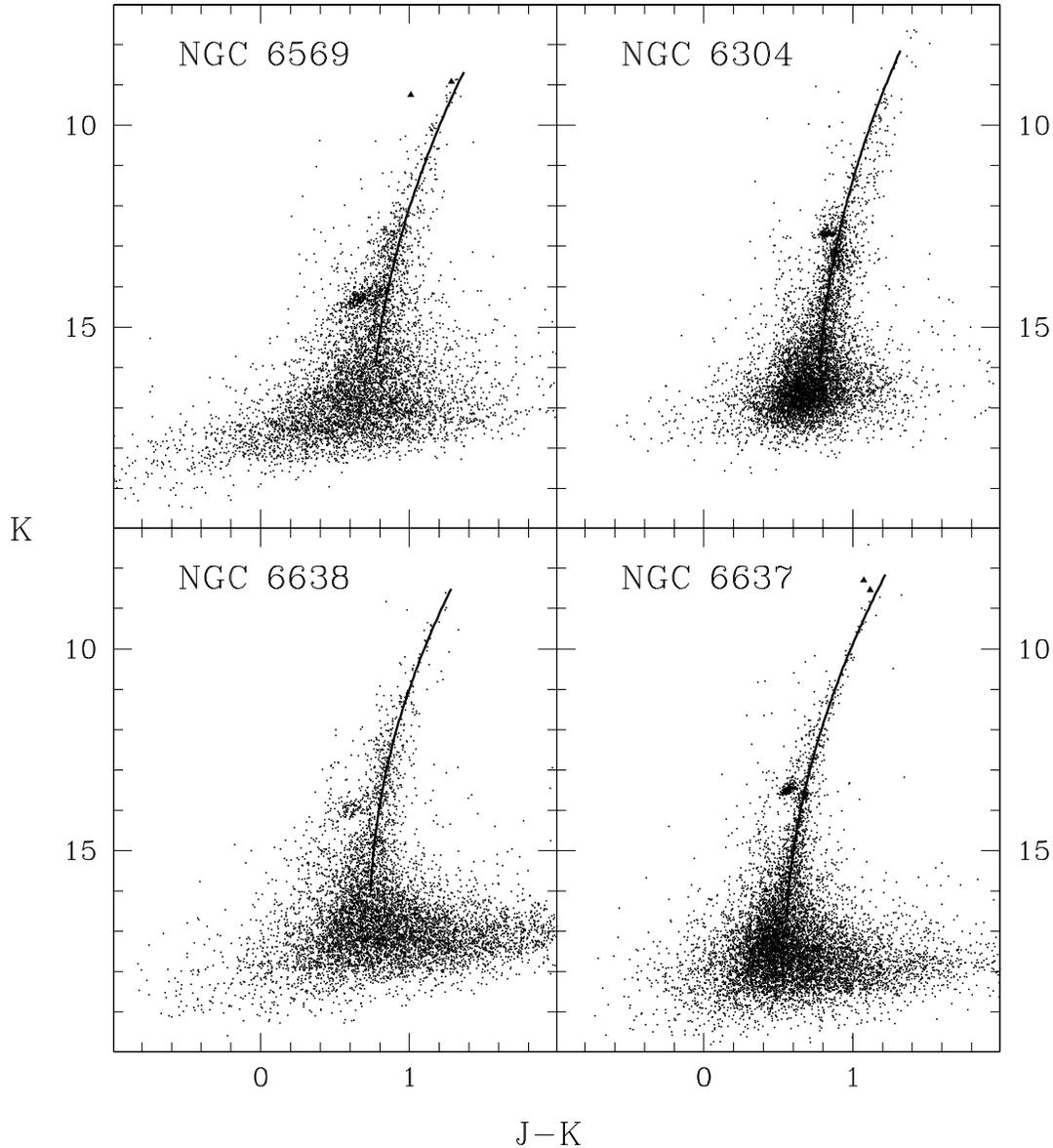}
\caption{K, J-K colour\--magnitude diagrams for the observed clusters. The
thick line in each panel indicates the RGB fiducial ridge line. 
Long period variables stars have been plotted as filled triangles.
Note that in the case of NGC~6304, because of bulge field contamination the RGB
fiducial ridge line is referring to the dominant blue component of the cluster
stellar population.} 
\label{kjk}
\end{figure*}

\begin{figure*}
\includegraphics[width=18cm]{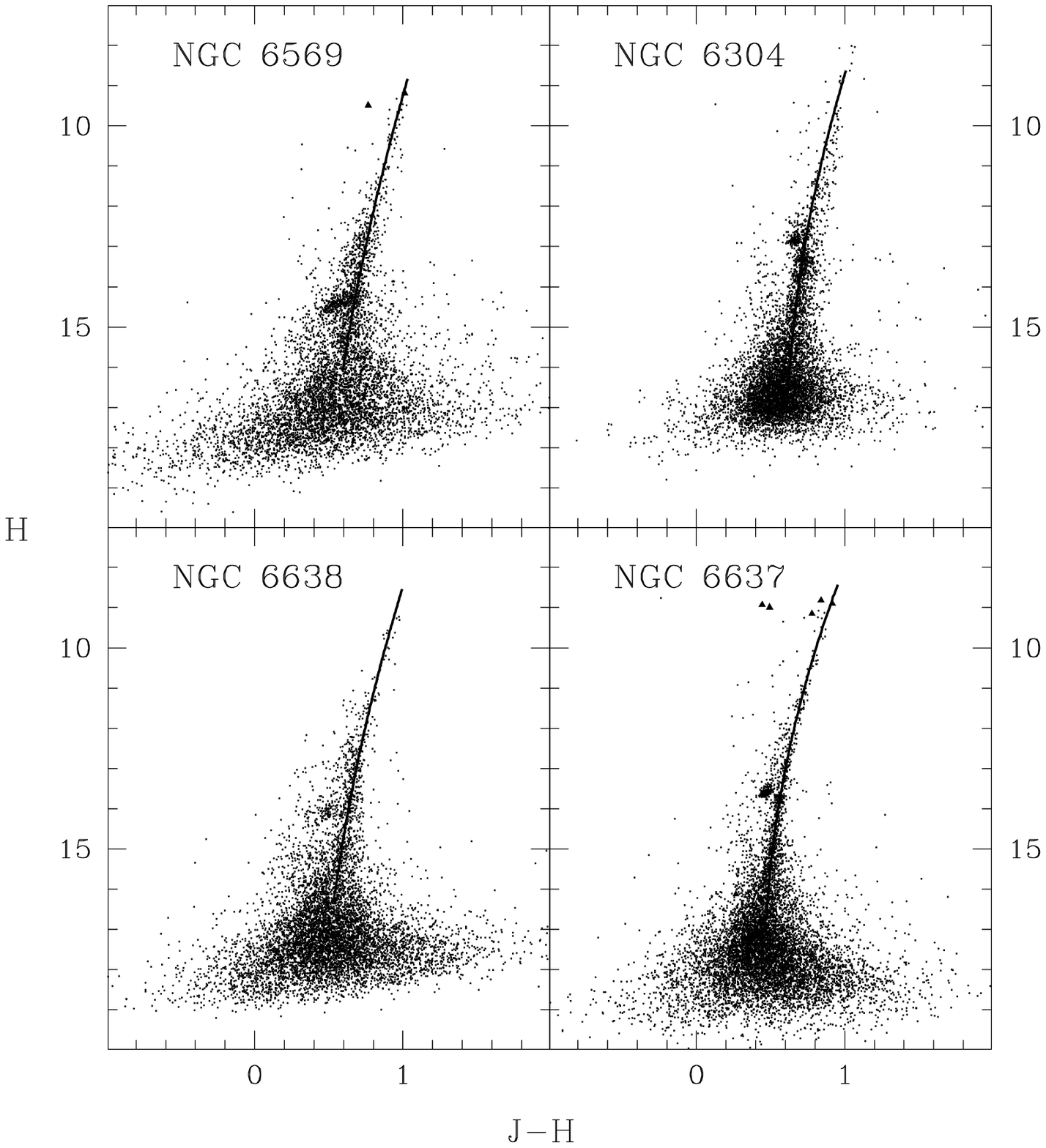}
\caption{As in Fig.~\ref{kjk} but for the H, J-H colours\--magnitude diagrams.}
\label{hjh}
\end{figure*}

In this framework, our group started a long\--term project devoted to fully
characterize the stellar populations in the Bulge GC system, by using 
Colour\--Magnitude Diagrams (CMDs) and Luminosity Functions (LFs) in the
near\--IR \citep[see][]{f00,v04,v04a,v04b,bb04}. 
As well known, in this spectral domain the reddening and
blanketing effects at high metallicity are much less severe than 
in the optical, moreover, 
the sensitivity to the physical parameters of cool stars is maximum and the
contrast between the red giants and the fainter MS population is greater than in
any other optical passband, drastically reducing the crowding, even in the
innermost core region \citep[see e.g.][ and references therein]{fer02}. 
As a first and major step forward in the study of the Bulge
stellar population we are performing an homogeneous survey of a
statistical significant sample of clusters in the near\--IR. 
Such a photometric screening of the Bulge GC system with similar accuracy  
as that obtained for the Halo GCs system \citep[see][]{f00}
will allow: {\it (i)} a global check of the stellar
evolution models in the high metallicity domain; 
{\it (ii)} a direct and quantitative comparison between the stellar
population in the Halo and Bulge GC system; 
{\it (iii)} an accurate calibration of
a few major integrated observables, to characterize the unresolved stellar
populations in extragalactic bulges. 

A set of near\--IR photometric indices (i.e. colours, magnitudes and slopes)
defined by \citet{f00} and widely discussed in 
\citet[][]{v04a,v04b} (hereafter {\it Paper~I} and {\it Paper~II}, 
respectively) have been 
derived to describe the morphology of the RGB together with its main 
evolutionary features, such the RGB Bump and Tip. 

In this paper we present a detailed IR study of the Red Giant Branch (RGB) 
sequence
for a sample of four metal\--rich Bulge GCs (namely, NGC~6304, NGC~6569, 
NGC~6637 and NGC~6638), affected by a moderate reddening $E(B-V)~<~0.6$
(see Table \ref{par}). 
These clusters have been subject of several studies, 
but mainly in the
optical range; no near\--IR CMDs are available for NGC~6638 
and NGC~6569. 
\citet{piotto02} published B,V HST photometry of all the programme clusters. The
derived CMDs clearly show an HB morphology and a curved RGB which are typical 
signatures of metal\--rich population; however only in the case of NGC~6637 
and NGC~6304 the Sub Giant Branch (SGB) and the 
Turn\--Off (TO) regions are well defined.
V, I ground\--based photometry of NGC~6304, NGC~6637 and
NGC~6638 has been presented by \citet{ros00} within a Galactic GCs survey 
devoted to derive the relative ages of the Milky Way GCs. However 
the relative high reddening and  
foreground contamination by MS stars, prevented a
sufficiently accurate detection of the TO level. Moreover, 
the RGB sequence is not properly sampled and defined. 
\citet{ort6304,ort6569} presented
V,I photometry of NGC~6304 and NGC~6569, respectively, 
finding the presence of a red clumpy HB, a curved 
RGB and high level of field contamination. 
The advantages of observing these clusters in the near\--IR has been
well demonstrated by \citet{f94} and
\citet{dav92} who published B,V,J,K photometry of NGC~6637 and 
V,K photometry of NGC~6304, respectively. 
The limited performances of the previous generation of IR arrays allowed 
only a partial sampling (both in space and deepness) of the RGB, which 
however is clearly defined in their observed K, V\--K CMDs,
thanks to the wide (V\--K) spectral 
baseline.

In the present study, for the programme clusters
we sampled the entire RGB extension in the J, H and K bands,
and we presented new estimates of the reddening
and distance by comparing the IR\--CMDs
with those of two reference clusters (47~tuc and M~107)
with well know extinction and distance modulus. 

The paper is organized as follow:
the observations and data reduction are presented in \S 2, while in \S 3 we
describe the properties of the CMDs. \S 4 describes the method
applied to derive an estimate of the cluster reddening and distance.
In \S 5 we present the study of the RGB sequence: by using LFs, CMDs and
RGB fiducial ridge lines, we derive 
the main RGB morphological and evolutionary features, and their
dependences on the cluster metallicity.
The results of the transformations between the observational and theoretical
planes for the RGB Bump and Tip luminosities are presented in \S 6. 
Finally, in \S 7 we briefly summarize our results.

\section{Observations and data reduction}
A set of J, H and K images of four Bulge clusters, 
namely NGC~6304, NGC~6637, NGC~6569 and NGC~6638,
was obtained at the European Southern 
Observatory (ESO), La Silla on June 2004, using the near\--IR camera
SOFI, mounted at the ESO/NTT telescope. During the observing runs two set of
data were secured:

{\it (i) Standard resolution set.} A series of images in the J, H and K bands
have been obtained by using SOFI in Large mode. In this combination the camera
has a pixel size of $0 \farcs 288$ and a total field of view of 
$4\farcm 9 \times 4\farcm 9$. The images are the combination of 42, 72, and 99
exposures each one 3\--sec long in the three pass-bands (J, H and K
respectively).

{\it (ii) High resolution set.} High resolution images of the inner region of
each cluster were also secured. The high resolution mode (SOFI coupled with
the focal elongator) yields a pixel size of $0 \farcs 146$ and a total field of
view $2\farcm 49 \times 2\farcm 49$. High resolution images are the average of
30 single exposures 1.2 sec\--long each.

All the secured images were roughly centered on the cluster center.
Note that the region covered by our observations allows us to sample 
a significant fraction of the total cluster light
(typically ${\sim}$80-95\%) in all the programme clusters. 
During the three nights of observation 
the average seeing was always quite good (FWHM${\approx}0\farcs 8$).
Every image has been background\--subtracted 
by using sky fields located several arcmin away from the cluster center, 
and flat\--field corrected 
using dome flat--fields, acquired with the standard SOFI calibration setup.

Standard crowded field photometry, including Point Spread Function
(PSF) modeling, was carried out on each 
frame by using DAOPHOTII/ALLSTAR \citep{dao94}. 
For each cluster, two photometric 
  catalogues (derived from high and standard resolution images), 
  listing the instrumental 
  J, H and K magnitudes, were obtained by cross-correlating the 
  single-band catalogues. 
  The standard and high resolution catalogues have been
  combined by means of a proper weighted average, weighting more the 
  high resolution measurements in the innermost region of the cluster.
  In principle, this strategy allows to minimize the blending effects. 
  The internal photometric accuracy 
  has been estimated from the rms frame\--to\--frame scatter of multiple
  stars measurements. Over most of the RGB extension, the internal errors are
  quite low (${\sigma}_J{\sim}{\sigma}_H{\sim}{\sigma}_K<0.03$ mag), increasing
  up to ${\sim}0.06$ mag at $K{\geq}16$.  
By using the Second Incremental Release Point Source Catalogue of 
2MASS, the instrumental magnitudes were then converted 
into the 2MASS photometric system.{\footnote {
An overall uncertainty of ${\pm}0.05$ mag in the zero\--point 
calibration in all three bands has been estimated. 
The observed cluster catalogues in the
2MASS photometric system are available in electronic form at the CDS.}
} 

The photometric catalogues have been also astrometrized onto 2MASS, 
using a procedure developed at the Bologna Observatory 
(P. Montegriffo, private communications) and successfully applied to 
other clusters 
\citep[see e.g.][{\it Paper~I} and II, and reference therein]{fer03,bb04}
providing rms residuals of $\approx$0.2 arcsec in both R.A. and DEC.

%
\section{Colour Magnitude Diagrams}
All stars detected over the entire science fields have been plotted
in the (K, J-K) and (H, J-H) CMDs shown in Fig.~\ref{kjk} and \ref{hjh}.
The main features of the CMDs are:

{\it i)} The Giant Branch is well populated in all clusters, 
even in the brightest magnitude bin, allowing a clear definition of the mean 
ridge line up to the Tip.

{\it ii)} The photometry is deep enough to reach the base of the RGB at
$K{\sim}H{\sim}15$ mag, i.e. $3-4$ mag below the Horizontal Branch (HB).
The SGB region is not well defined, preventing any 
feasible detection of the Turn\--Off level.

{\it iii)} The HB appears as a red clump well separated from the RGB. This is  
the typical morphology of 
high\--metallicity clusters such as 47~Tuc and M~107.

{\it iv)} In the CMDs of NGC~6304, the RGB is scattered,
and this is mainly due to contamination by bulge field giants 
 (see \S 4.1). 
The blue sequence at
$\rm (J-K){\approx}0.55$, extending up to $\rm K{\sim}14$ mag, corresponds to 
a disk MS contamination \citep[see][]{ort6304}.  

{\it v)} Long Period variables have been identified, 
to properly locate the RGB Tip (see \S~5.2).

\begin{table}
\begin{center}
\caption{The observed sample of clusters parameters from \citet{harris} 
catalogue.}
\label{sample}
\begin{tabular}{rccrrc}
\hline
Name & ${\alpha}_{2000}$ & ${\delta}_{2000}$ & l$^{\circ}$ & b$^{\circ}$ &
M$_V$ \\
NGC& & & & & \\
\hline
& & & & & \\
6304 & $17^h14^m32.5^s$ & $-29^{\circ}27{\arcmin}44{\arcsec}$ & 
355.83 & 5.38 & -7.26\\
6569 & $18^h13^m38.9^s$ & $-31^{\circ}49{\arcmin}35{\arcsec}$ & 
0.48 & -6.68 & -7.83\\
6637 & $18^h31^m23.2^s$ & $-32^{\circ}20{\arcmin}53{\arcsec}$ & 
1.70 & -10.27 & -7.47\\
6638 & $18^h30^m56.2^s$ & $-25^{\circ}29{\arcmin}47{\arcsec}$ & 
7.90 & -7.15 & -6.78\\
& & & & & \\
\hline
\end{tabular}
\end{center}
\end{table}

\begin{table*}
\begin{center}
\caption{Metallicity, reddening and distance 
modulus of the programme clusters.}
\label{par}
\begin{tabular}{lcccccccc}
\hline
Name & [Fe/H]$_{Z85}^a$ &[Fe/H]$_{CG97}^b$ &[M/H] & E(B-V)$_{H96}$ &E(B-V)$_{S98}$
&E(B-V)$_{derived}$ & (m-M)$^{H96}_0$ & (m-M)$^{derived}_0$ \\
\hline
& & & & & & & & \\
NGC~6304& -0.59 & -0.68 & -0.55 & 0.53 & 0.52 & 0.58 & 13.90 & 13.88 \\
NGC~6569& -0.86 & -0.79 & -0.64 & 0.55 & 0.44 & 0.49 & 15.15 & 15.40 \\
NGC~6637& -0.59 & -0.68 & -0.55 & 0.16 & 0.17 & 0.14 & 14.78 & 14.87 \\
NGC~6638& -1.15 & -0.97 & -0.69 & 0.40 & 0.42 & 0.43 & 14.91 & 15.07 \\
& & & & & & & & \\
\hline
\multicolumn{8}{l}{Notes:}\\
\multicolumn{8}{l}{$^a$~ Metallicity in the \citet{z85} scale.}\\
\multicolumn{8}{l}{$^b$~ Metallicity in the CG97 scale,  
as computed from the \citet{z85}
estimates by using equation (7) of CG97 }\\
\multicolumn{8}{l}{and following the prescription of \citep{f99}.}
\end{tabular}
\end{center}
\end{table*}

\begin{figure}
\includegraphics[width=9cm]{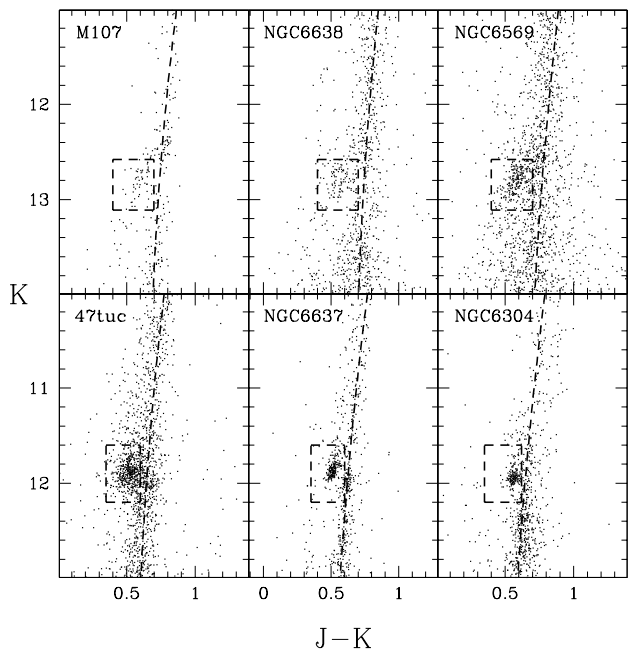}
\caption{A zoom of the RGB at the HB luminosity level in the K, J\--K plane
for the programme clusters. 47~Tuc is from {\it Paper~I}, M~107 
\citep[from][]{f00}. 
The observed RGB ridge lines are shown as dashed lines, while the dashed boxes
mark the HB region.} 
\label{hb}
\end{figure}
\section{Reddening and distance modulus}

As discussed in \S~2, the photometry of the four clusters presented here have
been reported in the 2MASS photometric system, hence they are fully consistent
with the 24 GCs published in {\it Paper~I} and {\it II}. This
allows to perform a detailed study of the RGB features by using the photometric
indices defined by \citet{f00} and calibrated in {\it Paper~I} and 
{\it Paper~II}.
Accordingly with the assumption done in those papers, here we adopted the
distance scale established by \citet{f99}, who derived the distance modulus for
a sample of 61 Galactic GCs based on an empirical estimation of the 
Zero Age Horizontal Branch level. Here the distance modulus
for the programme clusters is derived by 
comparing the CMDs with those of two reference 
clusters selected from the sample presented
by \citet{f00}, and applying a differential method.
As widely discussed in {\it Paper~I}, 
this procedure allows to 
derive simultaneously distance modulus and reddening estimates. 
As reference clusters we adopted 47~Tuc (for NGC~6304 and NGC~6637) 
and M~107 (for NGC~6569 and NGC~6638), since they have very similar  
metallicity and HB morphology, respectively (see Fig.\ref{hb}). 

The differences in colours and in magnitudes between the reference and the
programme clusters have been computed by shifting the CMD of each cluster to
match the RGB and the HB of the reference clusters. In particular, the colour
shifts ($\rm {\delta}$(J-H) and $\rm {\delta}$(J-K)) have been used to derive an
estimate of the relative reddening. The extinction coefficients listed by
\citet{SavMat} [A$_J$/E(B-V)=0.87, A$_H$/E(B-V)=0.54 and A$_K$/E(B-V)=0.38]
have been adopted. The results are listed in Table~\ref{par}, as can be seen our
reddening estimates nicely agree (within 0.1 mag) with those listed by
\citet{harris} and \citet{redS98}. The shifts in magnitudes 
($\rm {\delta}$J and $\rm {\delta}$K) once corrected for relative reddening have
been used to obtain an estimate of the distance modulus for each clusters.
The derived distances are listed 
in Table~\ref{par}, column [9] and compared with those found in the
\citet{harris} compilation (column [8]). Note that the distance modulus obtained
here for NGC~6637 turns out to be ${\sim}0.2$ mag larger than that found by
\citet{f94}. This is mainly due to a zero\--point difference
between the 2MASS photometric system and that
adopted by \citet{f94}.
\begin{figure}
\includegraphics[width=9cm]{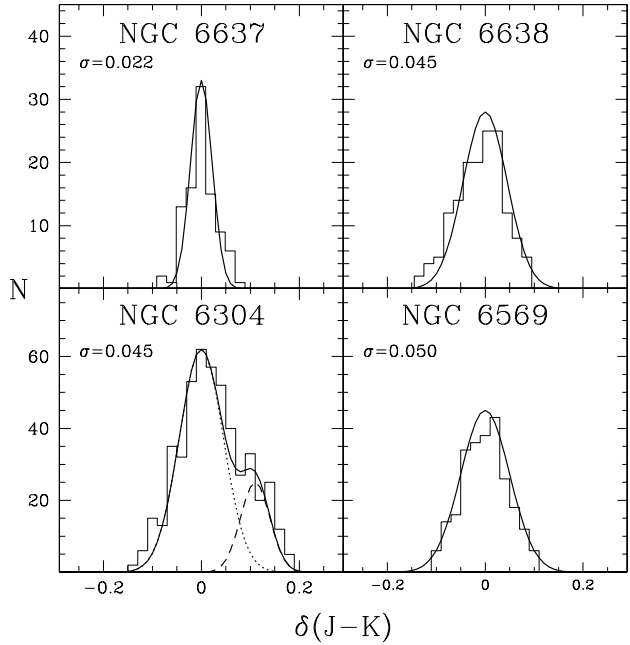}
\caption{J--K Colour distribution of the RGB stars in the 
$15.5<K<14$ magnitude range for NGC~6637 and NGC~6304 
and $12.5<K<11$ for NGC~6638 and NGC~6569. The solid thick lines are
the Gaussian best\--fits of the histograms.}
\label{hist}
\end{figure}
\subsection{The origin of the red sequence in NGC~6304}
The case of NGC~6304 deserves a brief discussion since a careful 
examination of its
CMDs has revealed the presence of a quite scattered RGB sequence   
(see Fig.~\ref{kjk} and ~\ref{hjh}).
The photometric errors are not large enough to justify such a spread in colour. 
Moreover the same RGB
morphology shown by our IR CMDs is also present in the 2MASS catalog of the same
region, ruling out the possibility that the observed feature is due to any
spurious effect in our photometry.

\begin{figure}
\includegraphics[width=9cm]{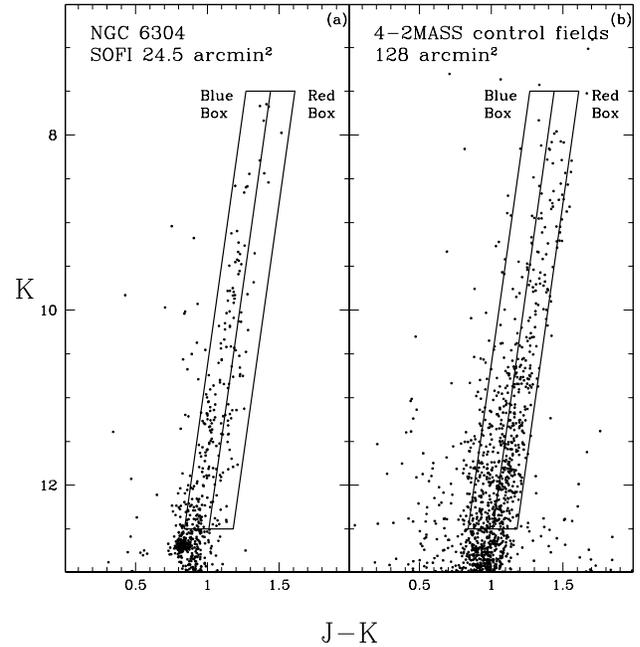}
\caption{K, J-K CMDs of RGB stars detected in the SOFI field (panel (a))
and by 2MASS in 4 control fields located ${\sim} 13 {\arcmin}$ from the 
cluster center
toward North, South, East and West (panel (b)).}
\label{6304field}
\end{figure}

\begin{figure}
\includegraphics[width=9cm]{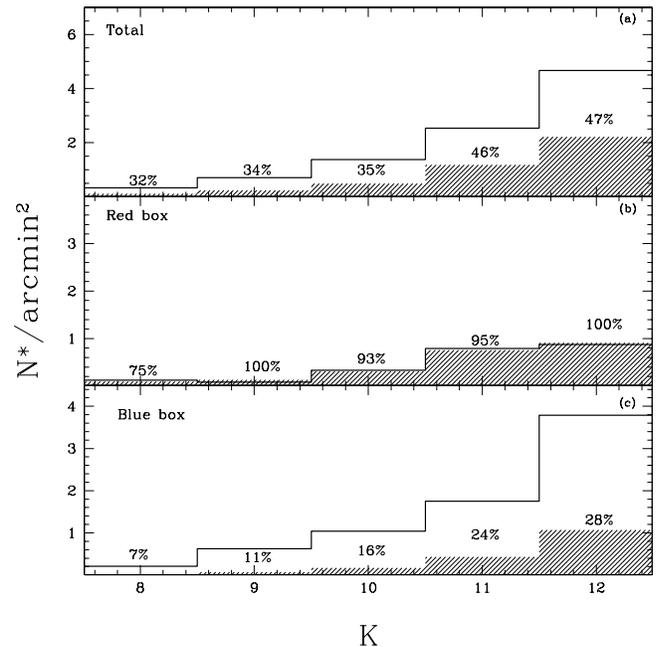}
\caption{Density distribution of stars detected in SOFI (open histograms)
and in 2MASS (shaded histograms) fields calculated within: both 
boxes (panel (a)),
only the Red Box (panel (b)) and only the Blue Box.
For each magnitude bin the percentage of field contamination is indicated. }
\label{6304cont}
\end{figure}

As can be clearly seen
in Fig.~\ref{hist}, in the case of NGC~6569, NGC~6637 and NGC~6638 the
(J\--K) colour distribution of the RGB stars, computed in a bin of ${\sim}$1.5 
mag, can be reproduced by a single Gaussian with a ${\sigma}$ value compatible
with the internal photometric errors. Conversely, NGC~6304 shows a
colour distribution with a pronounced tail toward the red.
At least two gaussian components are needed in order to reproduce the colour 
distribution shown in Fig.~\ref{hist}.
The main and bluer component is representative of the cluster RGB stars,
its colour distribution has in fact a ${\sigma}$=0.045 comparable with those of
the other clusters. A
secondary component at $\rm {\delta}$(J\--K)${\sim}$0.1 is needed in order to
fit the red tail of the colour distribution 
(see Fig.~\ref{hist}). 
In order to check whether the red sequence could be due to field contamination,
we investigated the field region around the cluster, by using 2MASS data. 
In particular, Fig.~\ref{6304field} shows the cluster CMD as obtained from our
observations (panel(a)), and for comparison the CMDs of 4 control fields
located at ${\sim} 13{\arcmin}$ from the cluster center (towards North, South,
East and West directions, panel (b)). As can be seen, most field giants lie on 
the red side of the NGC~6304 RGB. In order to make the effect more clear we 
divided the RGB region of the CMDs in two boxes: {\it Blue} and {\it Red Box}, 
respectively. 
As shown in the Fig.~\ref{6304field}, the vast majority of the field RGB 
stars lie in
the {\it Red Box}, while only few field giants are in the blue side of the box. 
To quantify the observed effect the number of field stars, per square arcmin, 
lying in the {\it Blue} and {\it Red Box} in bin of 1 mag has been computed. 
The derived density
distribution is compared with the same quantity in the SOFI field. The result
shown in Fig.~\ref{6304cont},
clearly indicates that almost all stars lying in the {\it Red Box} of panel (a) 
of Fig.~\ref{6304field} are due to Bulge field contamination (see i.e. panel (b)
of Fig.~\ref{6304cont}).

Note that the same procedure has been applied to all the programme clusters.  
In NGC~6569, NGC~6637 and NGC~6638 the level of bulge field
contamination is always less than 15\%.


\section{The main RGB features}
As already discussed in detail by \citet[][]{f00} and {\it Paper~I} and ~II, a
complete characterization of the RGB sequence, as a function of the cluster
metallicity, requires an appropriate definition of its morphological 
and evolutionary features, by using suitable CMDs and LFs. 
Briefly, the morphological characteristics of the RGB sequence can be obtained 
by computing a set of photometric indices (see \S 5.1)
such as {\it (i)} the colours at fixed magnitudes; {\it (ii)} the magnitudes 
at constant colours and {\it (iii)} the RGB slope.
In fact,   
{\it (1)} the absolute colour position, and {\it (2)} the morphology 
of the RGB sequence progressively change with increasing
cluster metallicity.
The Bump and Tip are the main RGB evolutionary features: the RGB Bump traces
a significant change in the evolutionary rate of the stars along the RGB, and 
the RGB Tip flags the end of the RGB evolutionary phase (see \S 5.3). 

As done in our previous studies, 
both the [Fe/H] in the \citet[][hereafter CG97]{cg97} scale and 
the [M/H] global metallicity, 
which takes into account the contribution of the 
${\alpha}$\--elements, are considered. 
The latter has been computed by using the equation  
\begin{equation}
[M/H]=[Fe/H]_{CG97}+log(0.638{\it f}_{\alpha}+0.362)
\end{equation}
where {\it f}$_{\alpha}$ is the 
enhancement factor of the ${\alpha}$\--elements 
(i.e. [$\rm \alpha/Fe]\approx+0.3~dex$]).
Following \citet{f00} and {\it Paper~I} and II, an 
${\alpha}$\--enhancement factor
linearly decreasing to zero for metal\--rich clusters with
[Fe/H]$_{CG97}<-1$ have been adopted.
However, it must be noted that,  
there is a growing number of recent high resolution spectroscopic observations 
of both Bulge cluster and field giants 
\citep[][]{rmw00,car01,ori02,ori04,zoc04,mwr04,ori05}
which point towards a constant value of
${\alpha}$\--enhancement up to solar metallicity.   
If a constant enhancement of $[\rm \alpha/Fe]\approx+0.3~dex$ 
over the full range of 
metallicity is adopted, the global metallicity of the programme clusters listed 
in Table~\ref{tabpar} must be increased, on average, by a few hundredths dex, 
the exact amount depending on their actual metallicity 
(the effect of such an 
assumption will be discussed in a forthcoming paper, Ferraro
et al. 2005, in preparation). 

\begin{table}
\begin{center}
\caption{Adopted parameters for the observed clusters.}
\label{tabpar}
\begin{tabular}{lcccc}
\hline
Name & [Fe/H]$_{CG97}$ & [M/H] & E(B-V) & (m-M)$_0$ \\
\hline
 & & & & \\
NGC~6304 & -0.68 & -0.55 & 0.58 & 13.88 \\
NGC~6569 & -0.79 & -0.64 & 0.49 & 15.40 \\
NGC~6637 & -0.68 & -0.55 & 0.14 & 14.87 \\
NGC~6638 & -0.97 & -0.69 & 0.43 & 15.07 \\
 & & & & \\
47~Tuc$^a$ & -0.70 & -0.59 & 0.04 & 13.32 \\
M~107$^a$ & -0.87 & -0.70 & 0.33 & 13.95 \\
 & & & & \\
\hline
\multicolumn{5}{l}{$^a$ From Table~2 of \citet{f99}.}
\end{tabular}
\end{center}
\end{table}
\subsection{The RGB morphology}
In order to derive the colours at fixed magnitudes and the
magnitudes at fixed colours the first step is the definition of the RGB fiducial
ridge lines from the CMDs.
We followed the prescriptions discussed in detail  
in \citet{f00,v04}. 
In the case of NGC~6304, where the
field contamination causes the observed splitting of the RGB sequence,
the fiducial ridge line has been computed by using only the main, blue 
component (see \S~4.1).

By using the values of reddening and distance 
modulus listed in Table~\ref{tabpar}, the observed RGB ridge lines of the
programme clusters have been also converted in the absolute plane.
Once obtained the intrinsic RGB ridge lines, a complete description of the RGB
morphology, as a function of the cluster metallicity, follows directly.

\begin{figure}
\includegraphics[width=9cm]{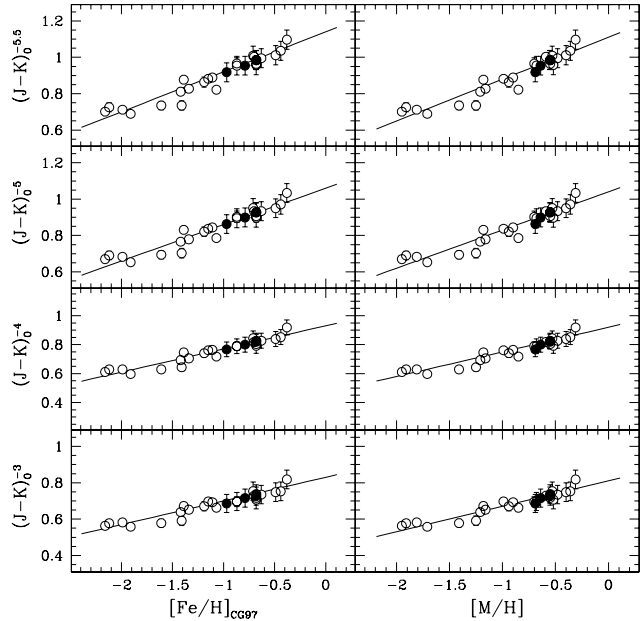}
\caption{RGB mean (J\--K)$_0$ colour at fixed M$_K$=(-5.5, -5, -4, -3)
magnitudes as a function of the CG97 metallicity scale (left panels) and 
of the global metallicity (right panels). 
Filled circles show the Bulge clusters
observed here, empty circles are the clusters presented in {\it Paper~I}. 
The solid lines are the calibration relations published in {\it Paper~I}.}
\label{coljk}
\end{figure}

\begin{figure}
\includegraphics[width=9cm]{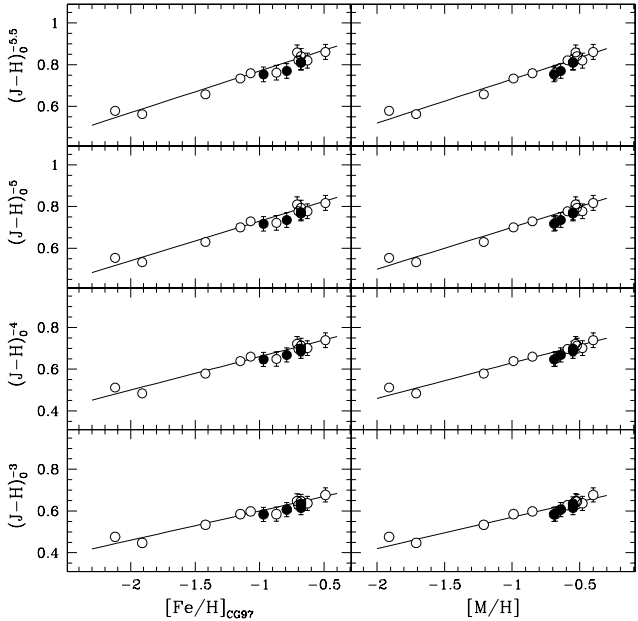}
\caption{The same as Fig.~\ref{coljk}, but for the (J\--H)$_0$ colours.}
\label{coljh}
\end{figure}

The RGB (J\--K)$_0$ and (J\--H)$_0$ intrinsic colours corresponding to 
four different
magnitude levels, namely M$_{K}$=$_{H}$ = (-5.5, -5, -4, -3), as defined in 
\citet{f00},
are listed in Table~\ref{inphot}, while their trends with 
the cluster metallicity are shown in Figs.~\ref{coljk} and 
~\ref{coljh}, respectively. 
Our results nicely agree with the
calibration relations (solid lines) derived in {\it Paper~I} 
(see equations A1\--A8 and A17\--A24 in the Appendix A of {\it Paper~I}).
As expected from previous studies \citep[][ an {\it Paper~I}]{cs95,f00,v04},
the linear scaling of the colours with the cluster metallicity is
confirmed, up to the highest metallicities. 
Table ~\ref{fe} and ~\ref{me} list the various photometric estimates 
of metallicity 
in both the CG97 and global metallicity scales, 
as computed using the relations (A1\--A8 and
A17\--A24) of {\it Paper~I} and the colours listed in Table ~\ref{inphot}.
They are all consistent within ${\pm}$0.1 dex. 

\begin{table*}
\begin{center}
\caption{Photometric indices describing the RGB location in colour and in
magnitude in the K,J\--K and H, J\--H planes, for the observed clusters.}
\label{inphot}
\begin{tabular}{lcccc}
\hline
Name & NGC~6304 & NGC~6569 & NGC~6637 & NGC~6638 \\
\hline
(J\--K)$^{M_K=-5.5}_0$& 0.984${\pm}$0.052 &0.954${\pm}$0.052  
&0.986${\pm}$0.051 &0.918${\pm}$0.050 \\
(J\--K)$^{M_K=-5}_0$  & 0.928${\pm}$0.052 &0.899${\pm}$0.052 
&0.927${\pm}$0.051 &0.863${\pm}$0.050 \\
(J\--K)$^{M_K=-4}_0$  & 0.827${\pm}$0.052 &0.800${\pm}$0.052 
&0.820${\pm}$0.051 &0.767${\pm}$0.050 \\ 
(J\--K)$^{M_K=-3}_0$  & 0.738${\pm}$0.052 &0.716${\pm}$0.052 
&0.727${\pm}$0.051 &0.686${\pm}$0.050 \\
(J\--H)$^{M_H=-5.5}_0$ &0.812${\pm}$0.035 &0.770${\pm}$0.035 
&0.809${\pm}$0.034 &0.754${\pm}$0.034 \\
(J\--H)$^{M_H=-5}_0$ &0.772${\pm}$0.035 &0.735${\pm}$0.035 
&0.766${\pm}$0.034 &0.717${\pm}$0.034 \\
(J\--H)$^{M_H=-4}_0$ &0.699${\pm}$0.035  &0.668${\pm}$0.035 
&0.686${\pm}$0.034  &0.647${\pm}$0.034 \\
(J\--H)$^{M_H=-3}_0$ &0.634${\pm}$0.035 &0.607${\pm}$0.035 
&0.616${\pm}$0.034 &0.584${\pm}$0.034 \\
M$^{(J-K)_0=0.7}_K$ &-2.52${\pm}$0.04 &-2.79${\pm}$0.04 
& -2.68${\pm}$0.04 &-3.19${\pm}$0.04 \\
M$^{(J-H)_0=0.7}_H$ &-4.02${\pm}$0.04 &-4.49${\pm}$0.03 
&-4.19${\pm}$0.03 &-4.77${\pm}$0.03 \\
RGB$_{slope}$ &-0.094${\pm}$0.005 &-0.089${\pm}$0.003 
&-0.092${\pm}$0.003 &-0.087${\pm}$0.004 \\
& & & & \\
\hline
\end{tabular}
\end{center}
\end{table*}
%
\begin{figure}
\includegraphics[width=9cm]{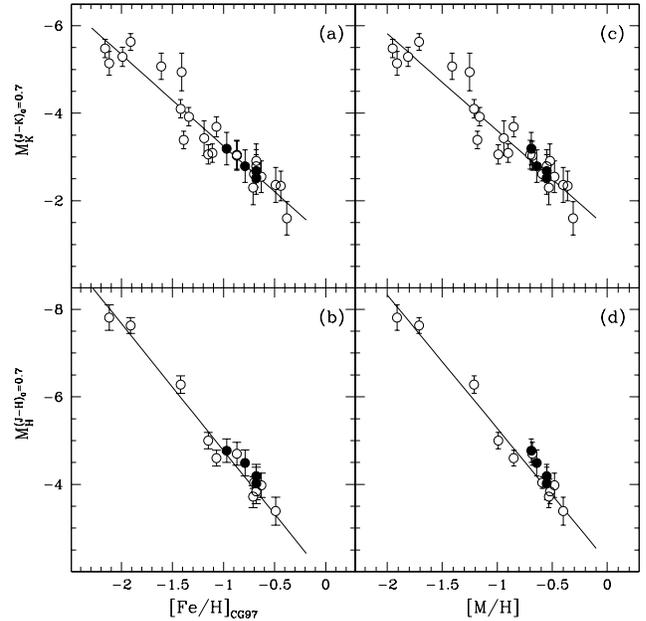}
\caption{Upper panels: M$_K$ at fixed (J\--K)$_0$=0.7 as a function of the
metallicity in the CG97 scale (a) and in the global scale (c). Lower panels:
M$_H$ at fixed (J\--H)$_0$=0.7 as a function of the
metallicity in the CG97 scale (b) and in the global scale (d). The filled
circles refer to the present sample, the empty circles the sample
of {\it Paper~I}. The solid lines are the calibration relations derived
in {\it Paper~I}.}
\label{mags}
\end{figure}

The behaviour of the RGB magnitudes at fixed 
colours as a function of metallicity has been investigated 
in all the observed clusters. 
Table~\ref{inphot} lists the derived M$_K$ and M$_H$ magnitudes at 
constant (J\--K)$_0$ and (J\--H)$_0$ colours. 
Fig.~\ref{mags} shows how these two indices 
 linearly correlate with the metallicity 
in both adopted scales.
The {\it Paper~I} sample together with
the corresponding calibration relations are also plotted. A
detailed discussion on the errors associated to the derived colours and
magnitudes can be found in {\it Paper~I}. Here we just remind that, 
while
the accuracy on the derived colours at fixed magnitudes is mainly driven by
the uncertainty on the distance modulus, the errors associated on the derived
magnitudes at constant colours depend on both reddening and distance
uncertainty with almost the same weight.
Column [4] of Table~\ref{fe} and ~\ref{me} lists the photometric estimate of
[Fe/H] and [M/H], respectively, derived by averaging the metallicity computed 
by using the relations A33\&A37 and A35\&A39 of {\it Paper~I} and the  
absolute M$_K$ and M$_H$ magnitudes listed in Table~\ref{inphot}. 

Another photometric estimate of the cluster metallicity is provided by the RGB
slope, which turns out to be particularly powerful being both reddening and
distance independent. The RGB slope in the K, J\--K plane for the programme
clusters have been computed following the prescriptions of {\it Paper~I}.
Photometric estimates of the clusters metallicity in both adopted scales
have been obtained by using the derived RGB slope and the relations 
A41\&A42 of {\it Paper~I}.
The results, listed in column [5] of Table~\ref{fe} and ~\ref{me}
are fully consistent with those obtained by using the colour and magnitude
photometric indices, within ${\leq}$0.1 dex.

\begin{figure}
\includegraphics[width=9cm]{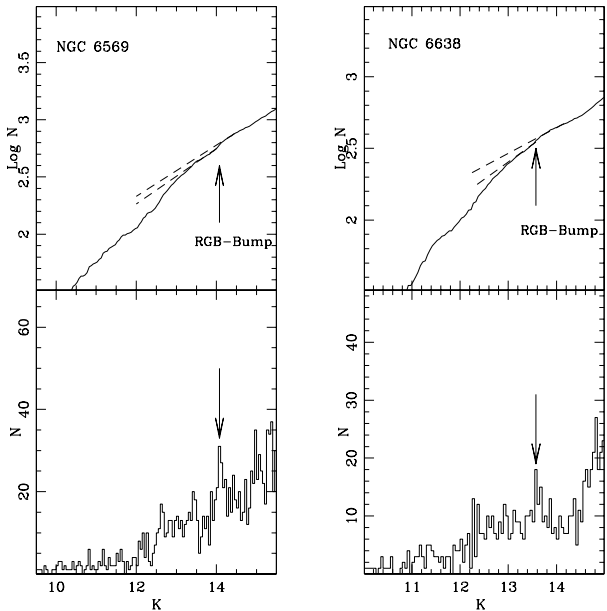}
\caption{Observed integrated (upper panels) and differential (lower panels)
luminosity functions for RGB stars in NGC~6569 and NGC~6638. The dashed lines
in the upper panels are the linear fits to the region above and below the RGB
bump.}
\label{bumpfl1}
\end{figure}

\begin{figure}
\includegraphics[width=9cm]{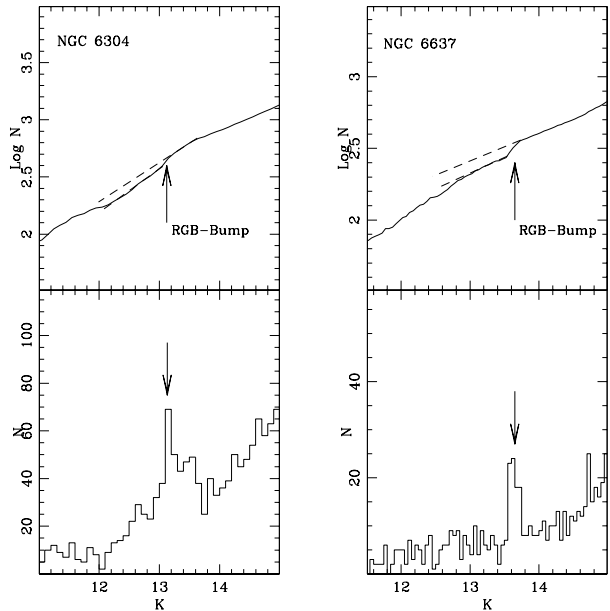}
\caption{As in Fig.~\ref{bumpfl1} but for NGC~6304 and NGC~6637.}
\label{bumpfl2}
\end{figure}

\begin{figure}
\includegraphics[width=9cm]{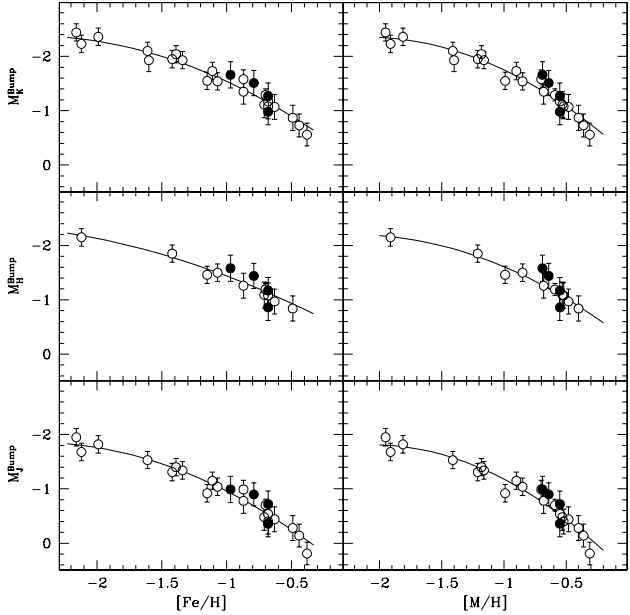}
\caption{The RGB Bump absolute K, H and J magnitudes as a function of the
cluster metallicity in the CG97 scale (left panels) and in the global scale
(right panels). Filled circles are the programme clusters, empty circles are
the {\it Paper~II} data set. The solid lines are the calibration relations from
{\it Paper~II}.}
\label{bump}
\end{figure}
\subsection{The RGB evolutionary features}

The RGB Bump flags the
point when, during the post\--MS evolution of low mass stars, the narrow
hydrogen\--burning shell reaches the discontinuity in the hydrogen distribution
profile, generated by the previous innermost penetration of the convective
envelope. The detection of the RGB Bump in Galactic and Local Group
stellar systems, has been subject of many studies 
\citep[see i.e.][ and {\it Paper~II}]
{fusi90,f99,manu99,f00,riello03,v04,bb04,mic01,mic02,lore02},
demonstrating how this feature can be safely identified by using 
suitable CMDs and LFs. 
In fact, the position of the RGB Bump corresponds to a peak in 
the differential LF and to a slope change in the integrated LF.
Following the same procedure adopted by \citet[][]{f00,v04} and {\it Paper~II}, 
the RGB Bump of the programme clusters has been detected in all the pass-bands.
As an example, Fig.~\ref{bumpfl1} and \ref{bumpfl2} show the  
integrated and differential LFs, in the K band, for RGB stars in all the
observed clusters.

The observed J, H and K Bump magnitudes are listed in Table \ref{inevo}.
Fig.~\ref{bump} shows the behaviour of the RGB Bump absolute magnitudes as a
function of metallicity in both the adopted scales.
The location of the Bump in NGC~6304 is somewhat uncertain due to 
the bulge field contamination (see \S~4.1). 
Indeed, the RGB\--LF in this cluster shows a few, 
broad/asymmetric peaks,   
the most pronounced being about 0.3 mag fainter than the 
corresponding peak in NGC~6637; by assuming the same age and same helium content
for the two clusters, this would imply a difference in metallicity of 
${\delta}$[M/H]${\sim}$0.15 dex.

By using equations 1\--6 of {\it Paper~II} linking the Bump 
magnitude to
the cluster metal content 
we have also derived independent 
estimates of metallicity in both the adopted scales (see Table~\ref{inevo}).
\begin{figure}
\includegraphics[width=9cm]{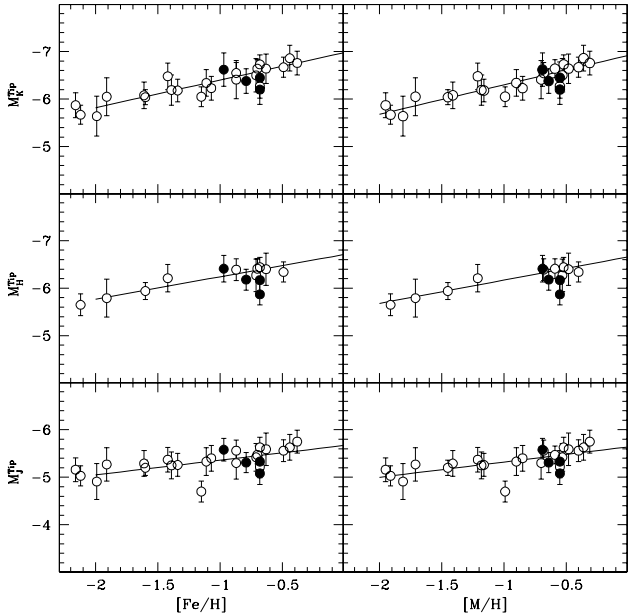}
\caption{The RGB Tip absolute K, H and J magnitudes as a function of the
cluster metallicity in the CG97 scale (left panels) and in the global scale
(right panels). Filled circles are the programme clusters, empty circles are
the {\it Paper~II} data set. The solid lines are the calibration relations from
{\it Paper~II}.}
\label{tip}
\end{figure}

For 
stellar population older than ${\tau}{\approx}$1\--2 Gyr (i.e. when stars less
massive than M${\approx}$2.0M$_{\odot}$ are evolving) the RGB Tip 
reaches its maximum luminosity, and it remains approximately constant with
increasing the population age. The calibration of the relation linking the RGB 
Tip
magnitude to the cluster metallicity is, therefore, a fundamental step in view
of using the RGB Tip luminosity as standard candle \citep[see][]{mic01}.
In {\it Paper~II} we recently published an updated calibration
of the RGB Tip magnitudes with varying metallicity in the near\--IR J, H and K
bands, based on a sample of 24 Galactic GCs. Here we present the estimates of
the RGB Tip luminosity for the programme clusters obtained following the same
procedure adopted there.
Briefly, this method is based on the assumption that the brightest
non\--variable star along the RGB can be considered as representative of the
RGB Tip level. 
The presence in our IR catalogue of variable stars along the RGB has been
checked by using the catalogue by \citet{variab}. 
The few variables are marked as
large filled triangles in the CMDs plotted in Fig~\ref{kjk} and ~\ref{hjh}.
The measures of the RGB Tip obtained for the programme clusters are listed in
Table~\ref{inevo}, while Fig.~\ref{tip}
shows the absolute RGB Tip magnitudes as a function
of the cluster metallicity in both the adopted scales.
As can be seen, our 
results nicely agree with the relations derived
in {\it Paper~II}.
Note that in the case of NGC~6637, the brightest non\--variable RGB 
star is different from that one identified by \citet{f00} as representative of
the Tip level.
This is due to the fact that the star used by \citet{f00} was a blend, which we
were able to separate into two stars, thanks to the superior  
spatial resolution of SOFI.

\begin{figure}
\includegraphics[width=9cm]{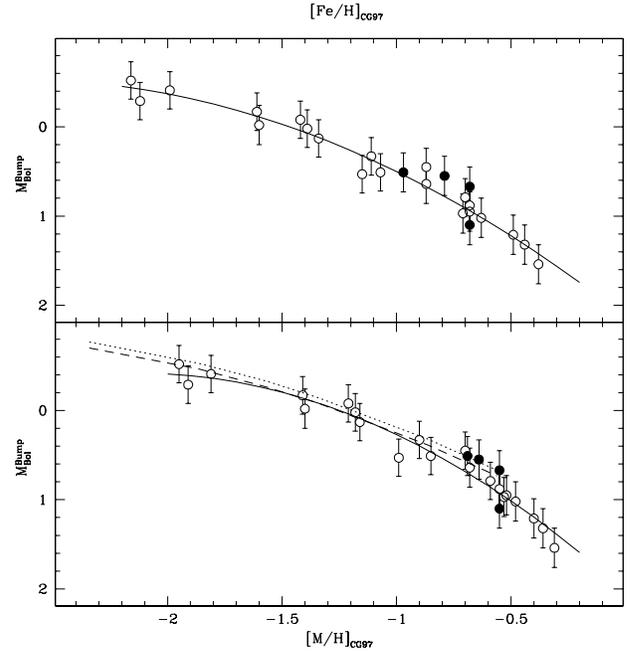}
\caption{Bolometric magnitudes of the RGB Bump as a function of the cluster
metallicity in both the CG97 (upper panel) and global (low panel) scales.
Filled circles show the present cluster sample and empty circles are the
{\it Paper~II} dataset. The solid lines are the calibration relations from 
{\it Paper~II},
the dashed and dotted lines (lower panel) are the theoretical prediction by
\citet{scl97} at t=12 and 14 Gyr, respectively.}
\label{bbol}
\end{figure}
\begin{figure}
\includegraphics[width=9cm]{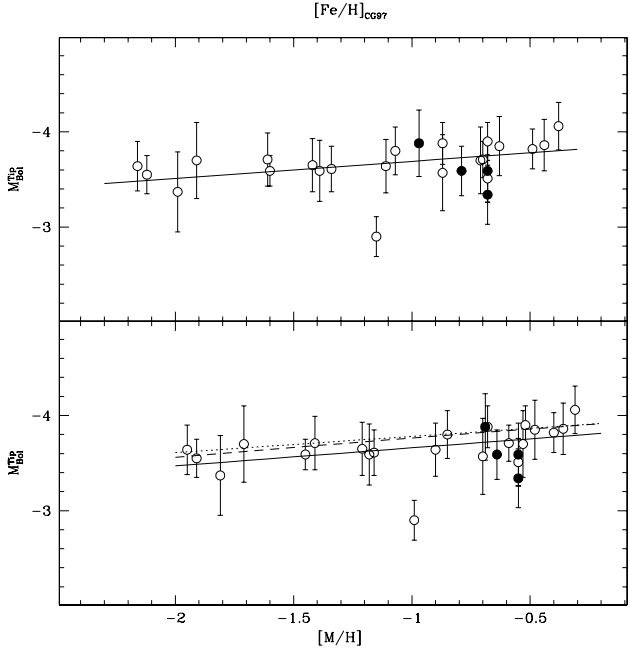}
\caption{Bolometric magnitudes of the RGB Tip as a function of the cluster
metallicity in both the CG97 (upper panel) and global (low panel) scales.
Filled circles show the present cluster sample and empty circles are the
{\it Paper~II} dataset. The solid lines are the calibration relations from 
{\it Paper~II}.
Two theoretical predictions have been plotted in the lower panel: \citet{caloi}
(dotted line) and \citet{sc97} (dashed line).}
\label{tbol}
\end{figure}
\section{The theoretical plane}
In order to compare the observed RGB evolutionary features with 
the model predictions, 
it is necessary to transform the observables into the theoretical plane, 
by converting the
absolute magnitudes into the bolometric one. In doing this,
the bolometric corrections for Population~II giants published
by \citet{paolo98} have been used. 
Table~\ref{inevo} lists the bolometric magnitudes of both
the RGB Bump and Tip for the observed cluster sample,
Figs.~\ref{bbol},\ref{tbol} show their trend with the cluster metallicity
in both the adopted scales. 
As can be seen, the values obtained for the programme clusters
well fit into the relations calibrated in {\it Paper~II}.

Theoretical predictions of the Bump features by \citet{scl97} 
show an excellent 
agreement with the observations (see Fig.~\ref{bbol}).
Because of the statistical fluctuations affecting the observed RGB Tip,
intrinsically poorly populated in GCs \citep{cast93},
the theoretical predictions 
of this feature \citep[see e.g.][]{caloi,sc97}
have to be considered as an upper luminosity 
boundary to the observed values.

\begin{table}
\begin{center}
\caption{Near\--IR and bolometric RGB Bump and Tip of the observed cluster
sample.}
\label{inevo}
\begin{tabular}{lcccc}
\hline
Name & NGC~6304 & NGC~6569 & NGC~6637 & NGC~6638 \\
\hline
 & & & & \\
J$_{Bump}$ &14.03${\pm}$0.10 &14.93${\pm}$0.05 &14.28${\pm}$0.05 &
14.45${\pm}$0.05 \\
H$_{Bump}$ &13.33${\pm}$0.10 &14.23${\pm}$0.05 &13.78${\pm}$0.05 &
13.73${\pm}$0.05 \\
K$_{Bump}$ &13.13${\pm}$0.10 &14.08${\pm}$0.05 &13.65${\pm}$0.05 &
13.58${\pm}$0.05 \\
 & & & & \\
J$_{Tip}$ &9.06${\pm}$0.23 &10.52${\pm}$0.21 &9.91${\pm}$0.23 &
8.61${\pm}$0.24 \\
H$_{Tip}$ &8.02${\pm}$0.25 &9.49${\pm}$0.22 &9.08${\pm}$0.22 &
8.89${\pm}$0.28 \\
K$_{Tip}$ &7.65${\pm}$0.33 &9.21${\pm}$0.26 &8.72${\pm}$0.31 &
8.61${\pm}$0.35 \\
 & & & & \\
M$^{Bump}_{bol}$ &1.10${\pm}$0.22 &0.55${\pm}$0.22 &0.67${\pm}$0.22 &
0.51${\pm}$0.22 \\
M$^{Tip}_{bol}$ &-3.59${\pm}0.33$ &-3.59${\pm}$0.26 &-3.34${\pm}$0.31 
&-3.88${\pm}$0.35 \\
\hline
\end{tabular}
\end{center}
\end{table}
\section{Summary and Conclusions}
High\--quality near\--IR photometry of NGC~6304, NGC~6569, 
NGC~6637 and NGC~6638 have been presented. 
By using CMDs and LFs we performed a detailed analysis
of the RGB morphology and evolutionary features. 
New estimates of the cluster
distance and extinction have been obtained.
A set of photometric indices has been used to provide 
photometric estimates of the cluster metallicity 
in both the [Fe/H] and [M/H] scales. We find [Fe/H]= -0.70, -0.88, -0.77, -1.00
and [M/H]= -0.54, -0.73, -0.62, -0.84 for NGC~6304, NGC~6569, NGC~6637 and
NGC~6638, respectively. 
The Bump and the Tip bolometric luminosities  
are in excellent agreement with the 
theoretical predictions.
\begin{table*}
\begin{center}
\caption{[Fe/H] photometric estimates for the observed clusters 
obtained by using the RGB morphological (colours, magnitudes and slope)
and evolutionary (Bump) features.}
\label{fe}
\begin{tabular}{lcccccc} 
\hline
Name & [Fe/H]$_{CG97}$ & [Fe/H]$_{col}$ & [Fe/H]$_{mag}$ & [Fe/H]$_{slope}$ &
[Fe/H]$_{Bump}$ & $<$[Fe/H]$>$\\
\hline
NGC~6304 & -0.68 & -0.73 & -0.69 & -0.71 & -0.55 & -0.70 \\
NGC~6569 & -0.79 & -0.90 & -0.84 & -0.82 & -0.97 & -0.88 \\
NGC~6637 & -0.68 & -0.78 & -0.76 & -0.76 & -0.77 & -0.77 \\
NGC~6638 & -0.97 & -1.06 & -0.98 & -0.87 & -1.11 & -1.00 \\
\hline
\end{tabular}
\end{center}
\end{table*} 

\begin{table*}
\begin{center}
\caption{[M/H] photometric estimates for the observed clusters  
obtained by using the RGB morphological (colours, magnitudes and slope)
and evolutionary (Bump) features.}
\label{me}
\begin{tabular}{lcccccc} 
\hline
Name & [M/H] & [M/H]$_{col}$ & [M/H]$_{mag}$ & [M/H]$_{slope}$ &
 [M/H]$_{Bump}$ & $<$[M/H]$>$\\
\hline
NGC~6304 & -0.55 & -0.58 & -0.56 & -0.57 & -0.44 & -0.54 \\
NGC~6569 & -0.64 & -0.74 & -0.70 & -0.68 & -0.80 & -0.73 \\
NGC~6637 & -0.55 & -0.62 & -0.62 & -0.61 & -0.61 & -0.62 \\
NGC~6638 & -0.69 & -0.90 & -0.83 & -0.72 & -0.92 & -0.84 \\
\hline
\end{tabular}
\end{center}
\end{table*} 

\section*{Acknowledgments}
The financial support by the Ministero
dell'Istruzione, Universit\'a e Ricerca (MIUR) is kindly acknowledged.
It is a pleasure to thank the anonymous Referee for a number of useful
comments which significantly improved the presentation of this work.
Part of the data analysis has been performed with the software developed by P.
Montegriffo at the Osservatorio Astronomico di Bologna (INAF).
We thank Sergio Ortolani for making available the optical photometry of
NGC~6569. We warmly thanks the ESO\--La Silla Observatory Staff for assistance 
during the observations.
This publication makes use of data products from the Two Micron All Sky Survey,
which is a joint project of the University of Massachusetts and Infrared
Processing and Analysis Center/California Institute of Technology, founded by
the National Aeronautics and Space Administration and the National Science
Foundation.

\label{lastpage}

\end{document}